\documentclass[pra,letterpaper,showpacs,preprint,superscriptaddress]{revtex4}

\usepackage{graphicx}
\usepackage[usenames]{color}

\begin{document}

\title{Spontaneous emission of a photon: wave packet structures and atom-photon
entanglement}
\author{M.V. Fedorov}
\affiliation{{A.M. Prokhorov General Physics Institute,
Russian Academy of Sciences, 38 Vavilov st, Moscow, 119991 Russia}}
\author{M.A. Efremov}
\affiliation{{A.M. Prokhorov General Physics Institute,
Russian Academy of Sciences, 38 Vavilov st, Moscow, 119991 Russia}}
\author{A.E. Kazakov}
\affiliation{{A.M. Prokhorov General Physics Institute,
Russian Academy of Sciences, 38 Vavilov st, Moscow, 119991 Russia}}
\author{K.W. Chan}
\affiliation{{Center for Quantum Information and Department of Physics and Astronomy,
University of Rochester, Rochester, NY 14627 USA}}
\author{C.K. Law}
\affiliation{Department of Physics, The Chinese University of Hong Kong, NT, Hong Kong SAR, China}
\author{J.H. Eberly}
\affiliation{{Center for Quantum Information and Department of Physics and Astronomy,
University of Rochester, Rochester, NY 14627 USA}}


\begin{abstract}
Spontaneous emission of a photon by an atom is described
theoretically in three dimensions with the initial wave function
of a finite-mass atom taken in the form of a finite-size wave
packet. Recoil and wave-packet spreading are taken into account.
The total atom-photon wave function is found in the momentum and
coordinate representations as the solution of an initial-value
problem. The atom-photon entanglement arising in such a process is
shown to be closely related to the structure of atom and photon
wave packets which can be measured in the coincidence and
single-particle schemes of measurements. Two predicted effects,
arising under the conditions of high entanglement, are anomalous
narrowing of the coincidence wave packets and, under different
conditions, anomalous broadening of the single-particle wave
packets. Fundamental symmetry relations between the photon and
atom single-particle and coincidence wave packet widths are
established.  The relationship with the famous scenario of
Einstein-Podolsky-Rosen is discussed.

\end{abstract}


\maketitle


\section{Introduction}
Beginning with the famous original derivation of natural linewidth
by Weisskopf and Wigner~\cite{WW}, spontaneous emission of atoms
has been considered traditionally, explicitly or not, in the
approximations of an infinitely heavy atom and an infinitely
narrow center-of-mass wave function. Rigorously, neither of these
approximations is ever correct. In a more realistic formulation,
for a finite-mass atom and a finite-size center-of-mass atomic
wave function, the problem of spontaneous emission was considered
by Rz\c{a}\.zewski and \.Zakowicz~\cite{RzaZak}. Chan et
al.~\cite{JHE} showed that such a formulation gives rise to
questions about photon-atom entanglement after emission. They
investigated this in one space dimension in the frame of Schmidt
mode analysis~\cite{JHE,JHE1,Kazik}. Atomic and photon
position-dependent Schmidt eigenfunctions were found numerically
and the conditions under which entanglement is large were
determined.

In this work we continue investigation of this problem. Its
solution is a bridge between two quite different regimes of
two-particle entanglement.  These two regimes are both concerned
with entanglement arising from the breakup of a composite object
into two ``fragments'' that are free to move away from the breakup
point, and in the ideal case are constrained only by momentum and
energy conservation.  One regime deals with zero-mass particles
(two photons), and here the most common context is spontaneous
parametric
down-conversion~\cite{Huang-Eberly93,Rubin,Monken-etal,BostonGrp,Barbosa,Law-etal00,Law-Eberly04,Kulik}.
The second regime treats two finite-mass fragments: electron and
ion in photoionization~\cite{PRA}, two atoms in molecular
dissociation~\cite{Opatrny-etal03,PRA,Opatrny-etal04,LANL}, or
even electron and positron in pair production~\cite{Krekora-etal}.
Two-particle scattering provides another view of breakup
entanglement~\cite{Grobe-etal94,Law04,Mishima-etal04}.  The regime
where one fragment is a photon and the other is massive raises
issues that deserve separate attention.

In contrast to the approach taken in Refs.~\cite{JHE,JHE1,Kazik},
we consider this bridging regime in a more realistic 3D picture.
We obtain the entangled position-dependent atom-photon wave
function as the solution of an initial-value problem. Atomic
recoil is taken into account and the initial atomic wave function
is taken in the form of a finite-size wave packet. We investigate
the structure of the photon and atomic wave packets that arise
after emission of a photon. In analogy with our treatment of
entanglement in photoionization and photodissociation~\cite{PRA},
we focus on experimentally accessible quantities.  We introduce
the parameter $R$ given by the ratio of the positional wave packet
widths to be measured singly or in coincidence, thus incorporating
both conditioned and unconditioned schemes of registration.
Spreading of the atomic wave packet is shown to play a crucial
role for the time evolution of the ratio $R$. Two specific
predicted effects are entanglement-induced narrowing of the
coincidence-scheme photon wave packet and, under different
conditions, a large broadening of the photon wave packet to be
found from the single-particle measurements. We note that
coincidence or conditional detection in regard to entanglement was
first carefully examined by Reid~\cite{Reid} in the context of
photonic squeezed states.  In that case one can define artificial
position and momentum variables using the photonic $\hat{a}$ and
$\hat{a}^\dagger$ operators, but there is not a simple analog of
wave packet spreading, which is our focus here.

In the next Section the general problem of the position-dependent
photon wave function is briefly discussed. In Section III we
outline the solution of the problem of spontaneous emission from a
finite-mass atom, characterized by a finite-size center-of-mass
wave packet. In Section IV we formulate a series of approximations
that are used to find an explicit expression for the photon-atom
wave function in the coordinate representation, and in section V
its entanglement features are analyzed. In Section VI we
investigate the wave packet structure in the coordinate
representation. In Section VII we introduce the main parameter $R$
and show that $R \gg 1$ characterizes the regions where an
anomalously narrow or wide wave packet can be found. Wave packets
in the momentum representation are analyzed in Section VIII, and a
series of symmetry relations between the wave packet widths is
derived. In Section IX we establish uncertainty relations
following from the high entanglement condition. One of them is the
coincidence uncertainty relation which is shown to restrict the
products of the particle's coordinate and momentum uncertainties
to be less than one (or $\hbar$). The relationship with the famous
Einstein-Podolsky-Rosen scenario~\cite{EPR} is described.
Experimental aspects are discussed briefly in Section X.

\section{Position-dependent photon wave function}

\noindent Although one can find examples~\cite{AB,LL} of a very
strongly formulated opinion that the position-dependent photon
wave function does not exist, in quantum optics the photon wave
function is accepted rather widely~\cite{MW,SZ,ON}. To outline the
way in which this concept can be introduced, let us consider the
state vector for the electromagnetic field, given by an arbitrary
superposition of one-photon states $\mid
1_{\vec{k}\lambda}\rangle$ with definite values of the wave vector
$\vec{k}$ and polarization $\lambda$
\begin{equation}
 \label{Psi-one-photon}
    |\Psi^{EM}(t)\rangle=\sum_{\vec{k}\lambda}C_{\vec{k}\lambda}\,
    e^{-i\omega_k t}\mid 1_{\vec{k}\lambda}\rangle\,,
\end{equation}
where $\omega_{k}=c|\vec{k}|$, $k_\alpha=n_\alpha 2\pi/L$ with
$\alpha=x,y, z$; $L$ is the normalization length, and
$\lambda=1,2$ is the polarization state of the photon. The state
vector (\ref{Psi-one-photon}) is conveniently normalized:
\begin{equation}
 \label{Psi-one-photon-normalization}
    \langle\Psi^{EM}(t)\mid\Psi^{EM}(t)\rangle=
    \sum_{\vec{k}\lambda}
    \mid C_{\vec{k}\lambda}\mid^{\,2}\,\equiv\sum_{\lambda}\int d\vec{k}\,
    \mid \widetilde{C}_{\vec{k}\lambda}\mid^{\,2}\,=1,
\end{equation}
where
$\widetilde{C}_{\vec{k}\lambda}=(L/2\pi)^{3/2}\,C_{\vec{k}\lambda}$.
Here and below we use in parallel both discrete and continuous
sets of wave vectors (in the normalization box $L^3$ and in free
space, respectively) with the sums and integrals over wave vectors
transformed from one to the other with the help of the relation
$\sum_{\vec{k}}=\frac{L^3}{(2\pi)^3} \int d\vec{k}$.

The expansion coefficients $C_{\vec{k}\lambda}$ in the
superposition (\ref{Psi-one-photon}) are interpreted naturally as
the photon wave function in the momentum representation. The
probability density to find (register) a photon with a momentum
around $\hbar\vec{k}$ is given by
\begin{equation}
 \label{prob-momentum}
 \frac{dw}{d(\hbar\vec{k})}=\frac{1}{\hbar^3}
 \sum_\lambda\left|\,\widetilde{C}_{\vec{k}\lambda}\right|^{\,2}.
\end{equation}

In quantum electrodynamics the operators of the field vector
potential and electric field-strength are given by
\begin{equation}
 \label{vector potential}
    \hat{\vec{\mathcal A}}(\vec{r})=\sum_{\vec{k}\lambda}
    \sqrt{\frac{2\pi\hbar c^2}{\omega_{k}\,L^3}}\,
    \left\{
    \vec{e}_{\vec{k}\,\lambda}\,
    e^{i\vec{k}\cdot\vec{r}}\,\hat{a}_{\vec{k}\lambda}+
    \vec{e}_{\vec{k}\,\lambda}^{\;*}\,
    e^{-i\vec{k}\cdot\vec{r}}\,\hat{a}_{\vec{k}\lambda}^{\dag}
    \right\}
\end{equation}
and
\begin{equation}
 \label{field-strength}
    \hat{\vec{\mathcal E}}^{\,}(\vec{r}^{\,})=i\sum_{\vec{k}\lambda}
    \sqrt{\frac{2\pi\hbar\omega_{k} }{L^3}}\,
    \left\{
    \vec{e}_{\vec{k}\,\lambda}\,
    e^{i\vec{k}\cdot\vec{r}}\,\hat{a}_{\vec{k}\lambda}-
    \vec{e}_{\vec{k}\,\lambda}^{\;*}\,
    e^{-i\vec{k}\cdot\vec{r}}\,\hat{a}_{\vec{k}\lambda}^{\dag}
    \right\}\,,
\end{equation}
where $\hat{a}_{\vec{k}\lambda}$ and
$\hat{a}_{\vec{k}\lambda}^{\dag}$ are the photon annihilation and
creation operators and ${\vec e}_{\vec{k}\,\lambda}$ are the
polarization vectors, ${\vec e}_{\vec{k}\,\lambda}\perp\vec{k}$.
The position-dependent density of the field energy is defined as
\begin{eqnarray}
    \label{energy-density-definition}
    \frac{d E_f(\vec{r},t)}{d\vec{r}}
    &=& \frac{1}{4\pi}\;
    \langle\Psi(t)|
    \hat{\vec{\mathcal{E}}}^{\,\dag} (\vec{r}) \cdot \hat{\vec{\mathcal{E}}} (\vec{r})
    |\Psi(t)\rangle \nonumber \\
    &=& \frac{1}{L^3} \left|
    \sum_{\vec{k}\lambda}
    \sqrt{\hbar\omega_{k}}\; \vec{e}_{\vec{k}\,\lambda}\,C_{\vec{k}\lambda}\,
    e^{i(\vec{k}\cdot\vec{r}-\omega_k t)}
    \right|^2 .
\end{eqnarray}
The total field energy of the state (\ref{Psi-one-photon}) is
given by the integrated density of energy of Eq.
(\ref{energy-density-definition})
\begin{equation}
 \label{en-total}
 E_f=\sum_{\vec{k}\lambda}\hbar\omega_{k}\,\left|\,C_{\vec{k}\lambda}\right|^{\,2}
 =\sum_\lambda\int
 d\vec{k}\;\hbar\omega_{k}\,\left|\,\widetilde{C}_{\vec{k}\lambda}\right|^{\,2}
 \equiv \hbar\overline{\omega},
\end{equation}
where, by definition, $\overline{\omega}$ is the mean photon
frequency of the superposition (\ref{Psi-one-photon}).

In general the position-dependent density of the field energy
$dE_f/d{\vec r}$ can also be presented in the form
\begin{equation}
 \label{energy-density-definition-ii}
   \frac{d E_f(\vec{r},t)}{d\vec{r}}=\hbar\overline{\omega}\left|{\vec\Psi}_{ph}({\vec
   r},t)\right|^2,
\end{equation}
where, following the Mandel-Wolf definition~\cite{MW},
${\vec\Psi}_{ph}({\vec r},t)$ is the vectorial position-dependent
one-photon photon wave function
\begin{eqnarray}
    \label{Psi-general}
    {\vec\Psi}_{ph}({\vec r},t)
    &=& \frac{1}{L^{3/2}\sqrt{\overline{\omega}}}\sum_{\vec{k}\lambda}
    \sqrt{\omega_{k}} \; \vec{e}_{\vec{k}\,\lambda}\;C_{\vec{k}\lambda}\,
    e^{i(\vec{k}\cdot\vec{r}-\omega_k t)}
    \nonumber \\
    &=& \frac{1}{(2\pi)^{3/2}\sqrt{\overline{\omega}}}
    \sum_{\lambda}\int d{\vec k}
    \,\sqrt{\omega_{k}}\; \vec{e}_{\vec{k}\lambda}
    \; \widetilde{C}_{\vec{k}\lambda}e^{i(\vec{k}\cdot\vec{r} - \omega_k t)}.
\end{eqnarray}
Owing to Eq.~(\ref{en-total}), the vectorial photon wave function
(\ref{Psi-general}) is also normalized:
\begin{equation}
 \label{norm}
 \int d\vec{r}\;\left|\vec{\Psi}_{ph}(\vec{r},t)\right|^2=1.
\end{equation}

It should be noted that the interpretation of
$\vec{\Psi}_{ph}({\vec r},t)$ of Eq.~(\ref{Psi-general}) as the
position-dependent photon wave function is only partially
satisfactory. By definition, this interpretation is good for the
calculation of the average photon energy or frequency
$\overline{\omega}$. But if we try to calculate with the help of
$\vec{\Psi}_{ph}({\vec r},t)$ the average values of other
quantities, e.g., the average photon momentum
$\hbar\overline{{\vec k}}$, in general the result will be wrong
because of the factor $\sqrt{\omega_{k}}$ in
Eq.~(\ref{Psi-general}). However, there is a class of states for
which this problem does not arise and the definition of the photon
wave function is non-controversial. This is the case of
narrow-band superpositions, for which the coefficients
$C_{\vec{k}\lambda}$ in~(\ref{Psi-one-photon}) are substantially
non-zero only inside a narrow spectral range $\Delta\omega$
\begin{equation}
    \label{narrow-band}
    |\, \omega_{k}-\overline{\omega}\, | \sim \Delta\omega\ll\overline{\omega}.
\end{equation}
In this case the factor $\sqrt{\omega_{k}}$ on the right-hand side
of Eq.~(\ref{Psi-general}) can be approximated by
$\sqrt{\overline{\omega}}$ to give simpler expressions for the
photon wave function
\begin{eqnarray}
    \label{Psi-coordinate}
    \vec{\Psi}_{ph}(\vec{r},t)
    &=& \frac{1}{L^{3/2}} \sum_{\vec{k}\lambda}\,C_{\vec{k}\lambda}\,
    \vec{e}_{\vec{k}\,\lambda}
    \, e^{i(\vec{k}\cdot\vec{r}-\omega_k t)}
    \nonumber \\
    &=& \frac{1}{(2\pi)^{3/2}}\int d\vec{k}
    \sum_\lambda\,\widetilde{C}_{\vec{k}\lambda}\,
    \vec{e}_{\vec{k}\,\lambda}\,
    e^{i(\vec{k}\cdot\vec{r}-\omega_k t)}.
\end{eqnarray}
In the case of spontaneous decay of atomic levels the spectral
width of the emitted light $\Delta\omega$ is the same as the decay
rate $\gamma$, which is always much less than the mean emitted
frequency $\overline{\omega}$. So, for spontaneously emitted
photon states the inequality (\ref{narrow-band}) is always
satisfied and the photon wave packet spectral width is relatively
small. The  Fourier transform (\ref{Psi-coordinate}) of the photon
momentum wave function establishes then an effective photon wave
packet width $c/\gamma$ in coordinate space.

Note that in speaking about the  position-dependent photon wave
function $\vec{\Psi}_{ph}(\vec{r},t)$ we assume that its squared
absolute value determines the probability density of the photon
registration by a detector located at the point $\vec{r}$.
Although below we will use the concept of the photon position
vector $\vec{r}_{ph}$, we will keep in mind that in fact this is
the position of the photon detector.

\section{Spontaneous emission}

\noindent To specify the problem to be considered, let us assume
that initially an atom is prepared in a pure excited $P$-state
with zero projection of its angular momentum upon the $z$-axis.
The preparation can be done, for example, with the help of
excitation from the ground $S$-state by a resonant laser
$\pi$-pulse with the linear polarization vector $\vec{e}_0$ along
the $z$ axis. If the pulse duration of the exciting pulse is short
compared to the life-time of the excited level, for spontaneous
emission the process of excitation is practically instantaneous,
and this reduces the problem of spontaneous emission to a solution
of an initial-value problem with a suddenly turned-on interaction.

Let the spontaneous emission of a photon arise from the atomic
transition back to the same ground state $|g\rangle$ from which
the atom was initially excited. Then, in the long-time limit $t\gg
1/\gamma$, where $\gamma$ is the decay rate, all the atomic
population returns to the ground state, and the two-particle
atom-photon state vector takes the form
\begin{equation}
    \label{state-vector}
    |\,\Psi\rangle= \sum_{\vec{q},\,\vec{k}}
    C_{\vec{q},\,\vec{k}\,}(t)
    \exp\left\{-i \left(\frac{ q^2}{2M}+E_g
    +\omega_{k}\right) t\right\}
    \,|\,g\rangle|\; \vec{q}\;\rangle|\,1_{\vec{k}\,}\rangle\,
    ,
\end{equation}
where $M$ is the total mass of the atom, $\vec{q}\,$ is its
momentum,  and $\vec{k}$ is the photon wave vector in a one-photon
state $|\,1_{\vec{k}\,}\rangle$.  Now and henceforth in this paper
we use a system of units with $\hbar=1$ and do not make any
difference between momenta and wave vectors of particles or
fields.

Multiplied by $(L/2\pi)^3$, the expansion coefficient
$C_{\vec{q},\,\vec{k}\,}(t)$ can be considered as the atom-photon
momentum-space wave function. It describes an entangled
atom-photon state when it is not factorable in the variables
$\vec{q}$ and $\vec{k}$. Summation over photon polarizations in
Eq.~(\ref{state-vector}) is unnecessary because for any given
$\vec{k}$ the atom can only emit a photon with the polarization
vector $\vec{e}_{k}$ in the plane $\{\vec{k},\vec{e}_0\}$
\begin{equation}
 \label{polar-vector}
 \vec{e}_{k}=\frac{k^2\vec{e}_0-\vec{k}(\vec{k}\cdot\vec{e}_0)}
 {k\sqrt{k^2-(\vec{k}\cdot\vec{e}_0)^2}}\;.
\end{equation}

The coefficients $C_{\vec{q},\,\vec{k}\,}(t)$ can be found in
Weisskopf-Wigner approximation to be given by~\cite{RzaZak}
\begin{equation}
    \label{Wigner-W}
    C^{(W-W)}_{\vec{q},\,\vec{k}\,}(t)\Big|_{\gamma t\gg 1} =
    - \frac{i e\,z_{eg}\,\omega_0\sqrt{2\pi}}
    {L^{3/2}\sqrt{\omega_{k}}}\,
    \frac{B(\vec{q}+\vec{k})\,\sin\vartheta_{k}}
    {\displaystyle\frac{q^2-(\vec{q}+\vec{k}\,)^2}{2M}
    +\omega_{k}-\omega_0+i\frac{\gamma}{2}}\,,
\end{equation}
where $z$ is the intra-atomic electron $z$-coordinate, $z_{eg}$ is
the matrix element of excited-ground state transition, and
$\vartheta_{k}$ is the angle between $\vec{k}$ and the
intra-atomic electron $z$-axis. The function $B(\vec{q})$ in
Eq.~(\ref{Wigner-W}) is the initial atomic wave function in the
momentum (wave vector) representation taken below in the Gaussian
form
\begin{equation}
    \label{Gaussian}
    B(\vec{q})
    =\left(\frac{2\pi}{L}\frac{a_0}{\sqrt{\pi}}\right)^{3/2}
    \exp\left(-\frac{a_0^2 \, q^2}{2}\right),
\end{equation}
so the corresponding initial atomic center-of-mass wave function
in the coordinate representation has a Gaussian form too
\begin{equation}
    \label{Gauss-initial-coord}
    \Psi_{at}(\vec{r}_{at},\,t=0)=\frac{1}{L^{3/2}}
    \sum_{\vec{q}}B(\vec{q})e^{i\vec{q}\cdot\vec{r}_{at}}
    =\left(\frac{1}{\sqrt{\pi} a_0}\right)^{3/2}
    \exp\left(-\frac{r_{at}^2}{2 a_0^2}\right).
\end{equation}
where $a_0$ is here not the Bohr radius but the initial size of
the atomic wave packet. Such a state of the center-of-mass motion
can be created, for example with the help of a trap. We assume
that at the same time $t=0$ when the trapped atom is excited and
spontaneous emission begins, the field of the trap is switched
off, and free spreading of the atomic center-of-mass wave packet
begins.

\section{The atom-photon wave function}

In analogy with the definition of the photon wave function in
Eq.~(\ref{Psi-general}), the position-dependent two-particle
atom-photon vectorial wave function can be defined as
\begin{equation}
    \label{at-phot-Psi}
    \vec{\Psi}(\vec{r}_{at},\vec{r}_{ph},t)
    =\frac{1}{L^3}\sum_{\vec{q},\,\vec{k}}
    C_{\vec{q},\,\vec{k}\,}(t)\,\vec{e}_{k}
    \,\exp\left\{i\left(\vec{q}\cdot\vec{r}_{at}+\vec{k}\cdot\vec{r}_{ph}\right)\right\}
    \,\exp\left\{-i\left(\frac{q^2}{2M}+E_g+\omega_k\right)t\right\},
\end{equation}
where $\vec{r}_{at}$ and $\vec{r}_{ph}$ are, correspondingly, the
atomic center-of-mass and photon position vectors, and
$\vec{e}_{k}$ is given by Eq.~(\ref{polar-vector}). In accordance
with the concluding remark of Section II, the terms of the
$``\text{position vectors}"$ are used conventionally. Rigorously,
the interpretation of the two-particle atom-photon wave function
is based on the assumption that its squared absolute value
determines the probability density of registering atoms and
photons by the corresponding detectors located at the points
$\vec{r}_{at}$ and $\vec{r}_{ph}$.

With $C_{\vec{q},\,\vec{k}\,}(t)$ and $B(\vec{q})$ taken from
Eqs.~(\ref{Wigner-W}) and~(\ref{Gaussian}) and with the
integration variable $\vec{q}$ replaced by $\vec{q}-\vec{k}$ we
can reduce Eq.~(\ref{at-phot-Psi}) to the form
\begin{eqnarray}
    \label{wf-1}
    \vec{\Psi}(\vec{r}_{at},\vec{r}_{ph},t)
    &=& -\displaystyle\frac{i\,e\,z_{eg}\,\omega_0 a_0^{3/2} \, e^{-iE_gt}}
    {(2\pi)^4\pi^{3/4}\sqrt{c}}
    \int d\vec{q}\,\exp\left[-\frac{q^2}{2}\left(a_0^2
    +\frac{it}{ M}\right) + i \,\vec{q}\cdot\vec{r}_{at}\right]
    \nonumber \\
    && \hspace{-2.5cm} \times \displaystyle\int_0^\infty k^{3/2}\,dk
    \,\exp\left(-\,\frac{i t}{2M}\,k^2\, \right)
    \int d\Omega_{k}\,\vec{e}_{k}\;
    \sin\vartheta_{k}\;
    \displaystyle\frac{\exp\left(i\,(\vec{k}\cdot\vec{\rho}-ckt)
    +\,\displaystyle\frac{it}{M}\,\vec{q}\cdot\vec{k}\,\right)}
    {-\displaystyle\frac{\vec{q}\cdot\vec{k}}{M}+\frac{ k^2}{2M}
    +\omega_{k}-\omega_0+i\frac{\gamma}{2}} ,
    \ \
\end{eqnarray}
where $\vec{\rho}=\vec{r}_{ph}-\vec{r}_{at}$ and $d\Omega_{k}$ is
a solid angle element in the direction of $\vec{k}$.

Analytical calculation of these integrals can be performed only
approximately. As the first approximation let us put
$\left|\vec{k}\right|\approx \omega_0/c$ in all the terms of the
integrand of Eq.~(\ref{wf-1}) proportional to $k^2$ and
$\vec{k}\cdot\vec{q}$. The precision of this approximation is
determined by small parameters proportional to $\omega_0/Mc^2\ll
1$. The second key approximation can be referred to as the far
zone approximation, which means that the distance $\rho$ is
assumed to be large enough, i.e., $k\rho\gg 1$. The validity of
this condition follows already from our original assumptions $t\gg
1/\gamma\gg 1/\omega$, which indicate immediately that at
$\rho\sim ct$ we have $k\rho\sim\omega t\gg 1$. In the far-zone
approximation the main contribution to the integral over
$d\Omega_{k}$ is given by those $\vec{k}$ close to the direction
of $\vec{\rho}$. Owing to this assumption we put
$\vec{k}\|\vec{\rho}$ everywhere in the integrand of
Eq.~(\ref{wf-1}) except in the factor
$\exp(i\vec{\rho}\cdot\vec{k})\equiv\exp(i\rho k x)$, where $x$ is
the cosine of the angle between $\vec{k}$ and $\vec{\rho}$. This
is the only remaining function of $x$ and it is easily integrated
to give two terms, proportional to $e^{ik\rho}$ and $e^{-ik\rho}$.
These two terms correspond to outgoing and incoming spherical
waves and, owing to the far-zone assumption, the incoming wave
gives an exponentially small contribution which can be dropped.
Thus, the result after the integration over $d\Omega_{k}$ is given
by
\begin{eqnarray}
    \label{wf-2}
    \vec{\Psi}(\vec{r}_{at},\vec{r}_{ph},t)
    &=& - \left[ \displaystyle \frac{e\,z_{eg}\,\omega_0 a_0^{3/2}
    \, e^{-iE_gt-i\omega_0^2t/2Mc^2}}
    {(2\pi)^3\pi^{3/4}\sqrt{c}} \right]
    \displaystyle \frac{\vec{e}_\rho^{\,\perp} \, \sin\vartheta_{\rho}}{\rho}
    \nonumber \\
    && \times
    \displaystyle\int d\vec{q}
    \;\exp\left[-\frac{q^2}{2}\left(a_0^2 + \frac{it}{M}\right)
    + i\,\vec{q}\cdot\vec{r}_{at}
    + \displaystyle\frac{it}{M}\,\frac{q_\rho\,\omega_0}{c}\,\right]\;
    \nonumber \\
    && \times
    \displaystyle\int_0^\infty dk \;
    \displaystyle\frac{k^{1/2}\;\exp\left[i\,k(\rho-ct)\right]}
    {-\displaystyle\frac{q_\rho\,\omega_0}{Mc}+\frac{\omega_0^2}{2Mc^2}
    +\omega_{k}-\omega_0+i\frac{\gamma}{2}} ,
\end{eqnarray}
where $\vec{e}_\rho^{\,\perp}$ is the unit vector perpendicular to
$\vec{\rho}$ and lying on the plane
$\{\vec{e}_0,\,\vec{\rho\,}\}$, $\vartheta_{\rho}$ is the angle
between $\vec{e}_0$ and $\vec{\rho}$, and $q_\rho$ is the
projection of the vector $\vec{q}$ on the direction of
$\vec{\rho}$.

Now, in accordance with the spirit of the Weisskopf-Wigner
approximation, the integral over $k$ is calculated by the residue
method with the lower limit of integration extended to $-\infty$
and the integrand continued analytically into the complex plane
$k$. Then we get
\begin{eqnarray}
    \label{residue}
    \vec{\Psi}(\vec{r}_{at},\vec{r}_{ph},t)
    &\Rightarrow&
    \displaystyle\frac{e\,z_{eg}\,\omega_0^{3/2}a_0^{3/2}}
    {(2\pi)^2\pi^{3/4}\ c^2}
    \left(\frac{\vec{e}_\rho^{\,\perp} \sin\vartheta_{\rho}}{\rho}\right)
    \,\theta(ct-\rho)\exp\left[\frac{\gamma}{2c}(\rho-ct)\right]
    \nonumber \\
    && \times \displaystyle\int d\vec{q}
    \; \exp\left[-\frac{q^2}{2}\left(a_0^2+\frac{it}{M}\right)+
    i\,\vec{q}\cdot\vec{R}\right],
\end{eqnarray}
where the symbol $``\Rightarrow"$ means that all the phase factors
independent of the integration variables are dropped and
\begin{equation}
    \label{cm-argument}
    \vec{R} \equiv \vec{r}_{at} + \frac{v_{rec}}{c}\,\vec{\rho}
    = \vec{r}_{at}\left(1-\frac{v_{rec}}{c}\right)+\frac{v_{rec}}{c}\,\vec{r}_{ph}.
\end{equation}
Integration over $d\vec{q}\,$ in Eq.~(\ref{residue}) can be easily
performed to give the following expression for the atom-photon
wave function:
\begin{eqnarray}
    \label{wf-last}
    \vec{\Psi}(\vec{r}_{at},\vec{r}_{ph},t)
    &\Rightarrow&
    \displaystyle\frac{e\,z_{eg}\,\omega_0^{3/2}}{\sqrt{2\pi}\;\pi^{3/4}\,c^2}
\left(\frac{\vec{e}_\rho^{\,\perp}
\sin\vartheta_{\rho}}{\rho}\right)\theta(ct-\rho)
    \exp\left[\frac{\gamma}{2c}(\rho-ct)\right]
    \nonumber \\
    && \times\displaystyle\frac{1}
    {\left(a_0 + it / Ma_0 \right)^{3/2}}
    \, \exp\left\{-\frac{\vec{R}^2}{2\left(a_0^2 + it/M\right)} \right\}\,,
\end{eqnarray}
where $v_{rec}=\omega_0/Mc$ is the atomic recoil velocity due to
the emission of a photon with momentum $\omega_0/c$. The squared
absolute joint wave function can then be written in the form
\begin{equation}
    \label{product}
    \left|\vec{\Psi}(\vec{r}_{at},\vec{r}_{ph},t)\right|^{\,2}
    = \left|\Psi_{rel}(\vec{\rho},\,t)\right|^{\,2}
    \times\left|\Psi_{cm}(\vec{R},\,t)\right|^{\,2}.
\end{equation}
Here $\Psi_{rel}(\vec{\rho},\,t)$ is the relative motion wave
function which depends only on the relative-motion position vector
$\vec{\rho}=\vec{r}_{ph}-\vec{r}_{at}$.  It takes the form of an
entanglement-free photon wave function  $\Psi_{ph}^{(0)}$
\cite{SZ}
\begin{equation}
\label{ent-free wf}
    \left|\Psi_{rel}(\vec{\rho},\,t)\right|^2\equiv\left|\Psi_{ph}^{(0)}(\vec{\rho},\,t)\right|^2
    =\displaystyle\frac{3\gamma \sin^2\vartheta_\rho}{8\pi
    c\,\rho^2}\,\theta(ct-|\vec{\rho}|) \, \exp\left[\frac{\gamma}{c}
    (|\vec{\rho}|-ct)\right] .
\end{equation}
On the other hand, although there is no center of mass,
$\Psi_{cm}(\vec{R},\,t)$ can be recognized as an analog of the
center-of-mass wave function of two massive particles in
situations such as photodissociation and photoionization
\cite{PRA}. In the case of atom-photon decay
$\Psi_{cm}(\vec{R},\,t)$ has the form of an entanglement-free
spreading atomic center-of-mass wave function $\Psi_{at}^{(0)}$.
Together with the initial condition given by
Eq.~(\ref{Gauss-initial-coord}), it reads
\begin{equation}
\label{ent-free wf-2}
    \left|\Psi_{cm}(\vec{R},\,t)\right|^2= \left|\Psi_{at}^{(0)}(\vec{R},\,t)\right|^2
    =\left(\frac{1}{\sqrt{\pi}a(t)}\right)^3
    \exp\left(-\frac{|\vec{R}|^2}{a^2(t)}\right),
\end{equation}
where $a(t)$ is the time-dependent width of the spreading atomic
wave packet
\begin{equation}
    \label{spreading}
    a(t) \equiv \left(a_0^2+\frac{t^2}{M^2a_0^2} \right)^{1/2} .
\end{equation}
We assume here that the atomic wave packet's spreading time
$t_{spr}\sim Ma_0^2$ is much longer than the atomic decay time
$\gamma^{-1}$.  This is true if the initial size of the atomic
wave packet is not too small: $a_0\gg 1/\sqrt{\gamma M}\sim 10 \
\text{nm}$ for $\gamma\sim 10^{8}$ sec$^{-1}$ and $M\sim 10^4
m_e$. Under this assumption the instant of time when the wave
packet spreading begins can be identified with the time $t=0$, at
which the atom is excited and quickly freed from the trap, and at
which the spontaneous emission process begins.

\begin{figure}[tb]
    \centering\includegraphics[width=8cm]{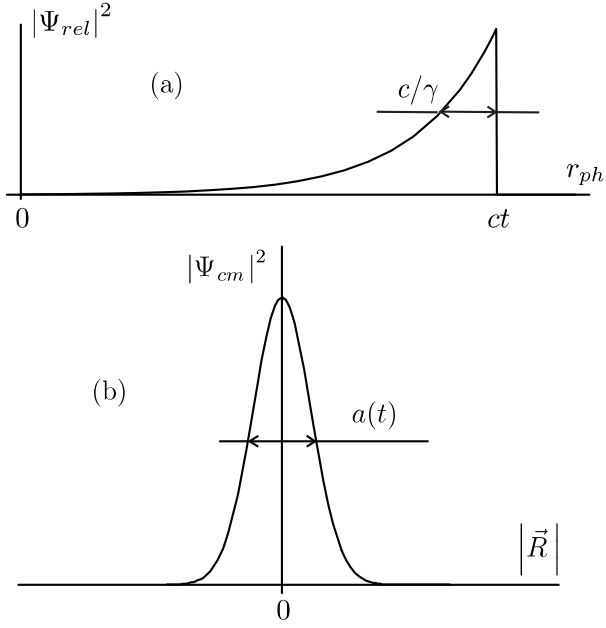}
    \caption{(a) The wave packet of a photon spontaneously
    emitted by an infinitely massive atom (Eq.~(\ref{ent-free wf})),
    and (b) the ``center-of-mass'' wave packet of a finite-mass ground-state atom
    (Eq.~(\ref{ent-free wf-2})).}
\label{Fig1}
\end{figure}
\begin{figure}[tb]
    \includegraphics[width=6cm]{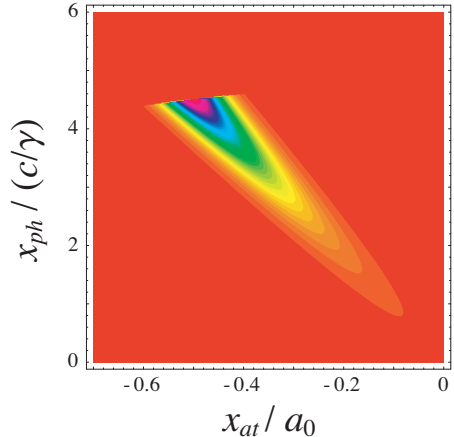}
    \caption{(Color online) The density plot of the one-dimensional
    analog of the total squared
    two-particle atom-photon wave function in Eq.~(\ref{product}) in the
    upper-left quadrant.
    Here $x_{at}$ and $x_{ph}$ denote the one-dimensional atom and photon coordinates.
    Parameters used are $\gamma t = 5, v_{rel}/c = 0.1$, and $ca_0/\gamma = 0.05$.}
\label{Fig2}
\end{figure}

The functions defined in Eqs.~(\ref{ent-free wf})
and~(\ref{ent-free wf-2}) are plotted in Fig.~\ref{Fig1}(a)
and~(b). They represent wave packets respectively for a photon
spontaneously emitted by a dot-like infinitely massive atom and
for a finite-mass ground-state atom with a spreading
center-of-mass wave function. Note that factorization of the total
wave function $\Psi$ into a product of the relative and
$``\text{cm}"$ parts is a rather general feature of the decaying
bipartite systems. It has been found to occur in the treatment of
photoionization and photodissociation~\cite{PRA} as well as, in a
wider sense, in spontaneous parametric down-conversion
\cite{Monken-etal,Law-etal00,Law-Eberly04, Kulik}. The product of
the functions shown in Fig.~\ref{Fig1} represents the total wave
function of Eq. ~(\ref{wf-last}), and in Fig.~\ref{Fig2} we show a
one-dimensional analog as a density plot. 
As it should be, the total atom-photon wave packet given by
Eq.~(\ref{product}) is appropriately normalized:
\begin{equation}
\label{norm2}
    \int d\vec{r}_{at} \int d\vec{r}_{ph}\;
    \left|\vec{\Psi}(\vec{r}_{at},\vec{r}_{ph},t)\right|^{\,2}
    \equiv \int d\vec{R}\int
    d\vec{\rho}\;\left|\vec{\Psi}(\vec{R},\vec{\rho},t)\right|^{\,2}
    = 1\,.
\end{equation}

\section{Entanglement}
As seen from Eqs.~(\ref{wf-last}) and (\ref{product}), the
atom-photon wave packet has the form of a product of two
$``\text{proto-packets}"$ given by Eqs.~(\ref{ent-free wf}) and
(\ref{ent-free wf-2}), related in this case to the wave packets of
a photon and an atom considered as independent particles. However,
the arguments of these proto-packets in the product in Eq.
(\ref{product}) are modified, or entangled, compared to those of
the independent single-particle wave packets of Eqs.
(\ref{ent-free wf}) and~(\ref{ent-free wf-2}). Each of the packets
$\left|\Psi_{rel}\right|^{\,2}$ and $\left|\Psi_{cm}\right|^{\,2}$
in Eq.~(\ref{product}) depends on both variables $\vec{r}_{at}$
and $\vec{r}_{ph}$, so that in these variables the total wave
function $\vec{\Psi}(\vec{r}_{ph},\vec{r}_{at},t)$ does not
factorize. This is the reason for, and clear indication of,
entanglement. It should be emphasized that the conditions for
obtaining atom-photon entanglement are finite mass of the atom and
finite size of its center-of-mass wave function. In the limits of
$M\rightarrow\infty$ and a $\delta$-localized atomic wave
function, the atom-photon wave function of Eq. (\ref{product})
factorizes, meaning that there is no entanglement
\begin{equation}
 \label{M->infty}
 \left|\vec{\Psi}(\vec{r}_{ph},\vec{r}_{at},t)\right|^{\,2}
 \rightarrow\left|\Psi_{ph}^{(0)}\left(\vec{r}_{ph},
 \,t\right)\right|^{\,2}\delta(\vec{r}_{at}).
\end{equation}

The main features of atom-photon entanglement are very similar to
those already described in photoionization and photodissociation
\cite{PRA}. The wave functions of fragments in these processes are
given by products of the fragments' center-of-mass and
relative-motion wave functions. Each of these proto-functions
depends on the position vectors of both fragments, and this is the
reason for entanglement. In the case of spontaneous emission, the
photon does not have a mass, and the decay product's center of
mass does not exist. Nevertheless, as shown above, it is possible
to present the total atom-photon wave function in the form of a
product of two proto-functions with arguments depending on both
atom and photon coordinates, and this explains entanglement in
atomic photoionization, molecular photodissociation, and
spontaneous emission from identical positions. Moreover, in
Eq.~(\ref{product}) the argument of $\Psi_{ph}$ equals the
difference of the photon and atomic position vectors
$\vec{r}_{ph}-\vec{r}_{at}$, which determines the same
relative-motion coordinates as in photoionization and
photodissociation.

As for the argument of $\Psi_{at}$ in Eq.~(\ref{product}), we
point out that its form is similar to the center-of-mass position
vector of massive particles, with the mass ratio substituted by
the ratio of velocities $v_{rec}/c$. Indeed, in the case of
photoionization the center-of-mass position vector can be written
as $\vec{r}_{cm}=\vec{r}_i(1-m_e/M)+\vec{r}_e(m_e/M)$ where
$\vec{r}_i$ and $\vec{r}_e$ are the ion and electron position
vectors, $m_e$ is the electron mass, and $M$ is the total mass of
the system $``{\rm electron‰}"$. By the substitutions
$\vec{r}_i\rightarrow\vec{r}_{at}$,
$\vec{r}_e\rightarrow\vec{r}_{ph}$, and $m_e/M\rightarrow
v_{rec}/c$ this expression reduces exactly to the argument of
$\Psi_{at}$ in Eq.~(\ref{product}). This makes the analogy between
the electron-ion and atom-photon entanglement almost complete.

Actually, in some sense the ratio of velocities is a more general
concept than the ratio of masses. Indeed, in the case of
photoionization the mass ratio can be presented in the form
$m_e/m_i=v_i/v_e$, where $v_i$ and $v_e$ are the magnitudes of the
classical ion and electron velocities after the break up of an
atom determined by the momentum conservation rule
$m_e\vec{v}_e+m_i\vec{v}_i=0$. Moreover, many results of of this
work and of Ref.~\cite{PRA} are valid also for the process of
downconversion~\cite{Law-Eberly04} (see sections VII and IX below)
with the velocity ratio substituted by one, because in this case
the break-up fragments are two photons, and their velocities are
equal.

\section{Position-dependent entangled atomic and photon wave packets}

One of the main ideas being presented here and in our earlier
works~\cite{PRA, Kazik} is the suggestion to use coincidence and
single-particle measurements of wave packets of particles arising
from decaying quantum systems for the analysis of particles'
entanglement and its manifestations. By single-particle
measurements we mean registration of one particle independent of
the other, e.g., detection of its position regardless of the
position of the other particle.  With repeated observations of
this kind, one can reconstruct the single-particle wave packet of
the chosen particle. As usual, the coincidence scheme requires two
detectors and registration of both particles. If the position of
one detector is kept constant and the position of another detector
is scanned, and if only joint signals from both detectors are
taken into account, such measurements can be used to reconstruct
the coincidence-scheme (i.e., conditional) wave packet of the
particle whose detector is scanned. The coincidence and
single-particle parameters of wave packets are indicated below by
the superscripts $^{(c)}$  and $^{(s)}$. Comparisons of the
coincidence and single-particle widths of wave packets provide
important information about entanglement and other correlation
measures of quantum systems undergoing breakup.

Mathematically, the coincidence and single-particle wave packets
are determined by appropriate conditional and unconditional
probability densities. In the case of spontaneous emission, for
example, for photon wave packets we have
\begin{equation}
 \label{Photon-abs}
  \frac{dw_{ph}^{(s)}(\vec{r}_{ph},t)}{d\vec{r}_{ph}}
  =\int d\vec{r}_{at}\left|\Psi(\vec{r}_{ph},\vec{r}_{at},t)\right|^2,
\end{equation}
and
\begin{equation}
 \label{Photon-cond}
  \frac{dw_{ph}^{(c)}(\vec{r}_{ph},t)}{d\vec{r}_{ph}}
  =\frac{\left|\Psi(\vec{r}_{ph}|\vec{r}_{at};\,t)\right|^2}
  {\int d\vec{r}_{ph}\left|\Psi(\vec{r}_{ph},\vec{r}_{at},t)\right|^2},
\end{equation}
where the second-place argument $|\vec{r}_{at}$ means $``\text{at
a given value of}\; \vec{r}_{at}"$. The same equations as
(\ref{Photon-abs}) and (\ref{Photon-cond}) with the substitution
$ph\rightleftharpoons at$ determine absolute (single) and
conditional (coincidence) probability densities for atomic wave
packets. In the following two subsections we shall calculate and
discuss the properties of the coincidence and single-particle
widths of the photon and atomic wave packets.

\subsection{Coincidence-scheme wave packets}

As shown previously, the square of the joint atom-photon wave
function in Eq.~(\ref{product}) is given by a product of two
proto-packets of Eqs. (\ref{ent-free wf}) and (\ref{ent-free
wf-2}) with entangled arguments. The photon proto-packet in
Eq.~(\ref{product}) depends on the difference of variables,
$\vec{r}_{ph}-\vec{r}_{at}$, and, as a function of this argument,
it has width equal to $c/\gamma$ (see Eq.~(\ref{ent-free wf}) and
Fig.~\ref{Fig1}(a)). As the entangled argument of the photon
proto-packet is simply a difference of variables, it is clear that
the same width $c/\gamma$ characterizes the photon proto-packet in
its dependence on either $\vec{r}_{ph}$ or $\vec{r}_{ph}$ at a
given value of the other variable.

On the other hand, the entangled argument of the atomic
proto-packet in Eq.~(\ref{product}) is more complicated,
approximately equal to
$\vec{r}_{at}+\left(v_{rec}/c\right)\vec{r}_{ph}$. As seen from
Eq.~(\ref{ent-free wf-2}) and Fig.~\ref{Fig1}(b), the width of the
atomic proto-packet with respect to its entangled argument equals
$a(t)$ as given in Eq. (\ref{spreading}). From here and from the
form of the entangled argument of the atomic proto-packet we
conclude immediately that its width with respect to $\vec{r}_{at}$
at a given value of $\vec{r}_{ph}$ is $a(t)$, is the same as that
of the unentangled atomic wave function of Eq. (\ref{ent-free
wf-2}), whereas the width of the atomic proto-function in its
dependence on $\vec{r}_{ph}$ at a given value of $\vec{r}_{at}$ is
equal to $a(t)\times (c/v_{rec})$.

\begin{figure}[tb]
    \includegraphics[width=6cm]{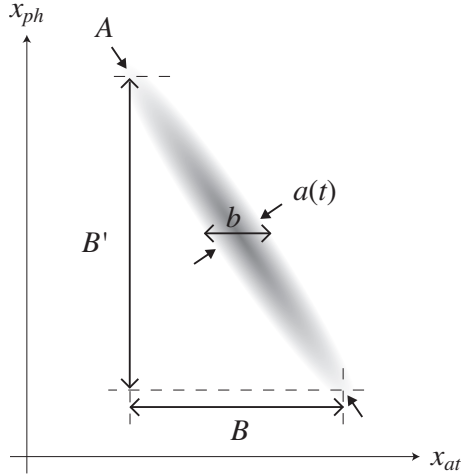}
    \caption{The density plot of the one-dimensional Gaussian model to the total
    squared two-particle atom-photon wave function in Eq.~(\ref{product}),
    with $x_{at}$ and $x_{ph}$ denoting the one-dimensional atom
    and photon coordinates.  Here $A=c/\gamma$ gives the width of the
    relative proto-packet $|\Psi_{rel}|^2$ in Eq.~(\ref{ent-free wf}) and
    $a(t)$ is the width of the CM proto-packet $|\Psi_{cm}|^2$
    in Eq.~(\ref{ent-free wf-2}).
    The coincidence width $\Delta x_{at}^{(c)}$ and the single width
    $\Delta x_{at}^{(s)}$ of the atom are given by $b$ and $B$ respectively,
    and $B'$ is the single-particle
    width $\Delta x_{ph}^{(s)}$ of the photon.}
\label{Fig3}
\end{figure}

The widths of the two-particle wave packet of Eq. (\ref{product})
in its dependence on either $\vec{r}_{ph}$ at a given
$\vec{r}_{at}$ or on $\vec{r}_{at}$ at a given $\vec{r}_{ph}$ can
be envisioned by reference to Fig.~\ref{Fig2} or to the generic
plot in Fig.~\ref{Fig3}. Qualitatively, the widths are the minima
of the corresponding widths of the proto-packets in their product
in Eq. (\ref{product}), which gives the following expressions for
the coincidence (conditional) widths of the photon and atomic wave
packets
\begin{equation}
    \label{coincidence}
    \left.\Delta r_{ph}^{(c)}(t) \equiv \Delta r_{ph}\right|_{\vec{r}_{at}=const}
    \sim \min\left\{\frac{c}{\gamma}\, , \; \left(\frac{c}{v_{rec}}\right)a(t)\right\} ,
\end{equation}
and
\begin{equation}
    \label{at-coinc-width}
    \left.\Delta r_{at}^{(c)}(t) \equiv \Delta r_{at}\right|_{\vec{r}_{ph}=const}
    \sim \min\left\{\frac{c}{\gamma}\, , \; a(t)\right\} .
\end{equation}
These minima are conveniently expressed by introducing a control
parameter analogous to those used in previous
works~\cite{JHE,PRA}:
\begin{equation}
 \label{eta}
    \eta(t) = \frac{\gamma a(t)}{c} ,
\end{equation}
and in terms of $\eta(t)$ they take the forms
\begin{equation}
 \label{widths(eta)}
 \Delta r_{ph}^{(c)}(t)  \sim  \frac{c}{\gamma}\,\frac{\eta(t)}{\sqrt{\eta^2(t)+v_{rec}^2/c^2}} ,
\quad\text{and}\quad
 \Delta r_{at}^{(c)}(t) \sim  \frac{a(t)}{\sqrt{1+\eta^2(t)}} .
\end{equation}

Two notes to be made in connection with these and later
definitions of the wave packet widths concerns the precision of
such formulas. First, the rigorously defined variances can depend
on the shape of the wave packets. Second, in principle, the
coincidence widths of the wave packets can depend on the values of
the fixed variables, e.g., $\Delta r_{ph}^{(c)}$ can depend on
$\vec{r}_{at}$, etc. For these reasons Eqs.~(\ref{coincidence}),
(\ref{at-coinc-width}), (\ref{widths(eta)}) and similar ones
should be understood as giving estimates of widths maximised with
respect to the fixed variables and with undefined coefficients of
the order of one in front of expressions on the right-hand sides.
In a model of two 1D Gaussian entangled wave packets, as
demonstrated in Fig.~\ref{Fig3}, the relations (\ref{widths(eta)})
become exact, and the symbol $``\sim"$ can be replaced by $``="$.

As discussed above, $c/\gamma$ and $a(t)$ can be considered as the
natural photonic and atomic wave packet widths found under the
most often used assumptions: in the approximations of an
infinitely heavy atom with the $\delta$-localized center-of-mass
wave function for photonic wave packet, and in the case of a
finite-mass non-excited and non-emitting atom for its spreading
center-of-mass wave function. Any deviations from these natural
widths can be considered as manifestations of entanglement in the
two-particle system. To characterize these deviations, which are
discussed below, it is convenient to introduce relative
dimensionless widths
\begin{equation}
 \label{rel-coinc-ph}
  \delta r_{ph}^{(c)}(t)=\frac{\Delta
  r_{ph}^{(c)}(t)}{c/\gamma}\sim
  \frac{\eta(t)}{\sqrt{\eta^2(t)+v_{rec}^2/c^2}}
\end{equation}
and
\begin{equation}
 \label{rel-coinc-at}
  \delta r_{at}^{(c)}(t)=\frac{\Delta r_{at}^{(c)}(t)}{a(t)}\sim
  \frac{1}{\sqrt{1+\eta^2(t)}}.
\end{equation}
The $``\text{natural}"$ values of these relative widths in a
system without any entanglement are equal to one.

\subsection{Single-particle wave packet widths}

In accordance with the definition of Eq.~(\ref{Photon-abs}) the
single-particle wave packets are related to the integrated squared
absolute value of the two-particle wave function. There are two
ways of performing the integrations over $\vec{r}_{at}$ or
$\vec{r}_{ph}$ analytically. First, we can model the photon
proto-packet in Eq. (\ref{ent-free wf}) by a Gaussian one. Then
Eq.~(\ref{product}) takes the form of a product of two Gaussian
packets with entangled variables and integration is carried out
easily. We will use this method below in an analysis of the
momentum-space wave packet. Here we will use another approach
based on the consideration of two opposite cases, when one of two
proto-packets in the product on the right-hand side of Eq.
(\ref{product}) is much narrower than the other one. Then the
integrations can be carried out approximately, and the approximate
expressions for the integrated absolute single-particle
probability densities can be used for evaluation of the
single-particle widths of wave packets.

In the case of a photon single-particle wave packet the squared
absolute value of the two-particle wave function must be
integrated over $\vec{r}_{at}$. Because of the forms of the
entangled arguments of the atomic and photon proto-packets, in the
case of integration over $\vec{r}_{at}$ the limits of relatively
narrow and wide atomic proto-packets are separated by the
conditions of small and large values of the control parameter
$\eta(t)$ as defined in Eq.~(\ref{eta}). With these conditions
being used, the result of the integration takes the form
\begin{equation}
    \label{single-w-f}
    \frac{dw^{(s)}}{d\vec{r}_{ph}}=\int d\vec{r}_{at}\,
    \left|\vec{\Psi}(\vec{r}_{at},\vec{r}_{ph},t)\right|^{\,2}
    = \left\{
    \begin{array}{cl}
    \left|\Psi_{cm}(\vec{r}_{ph})\right|^{\,2}\,, &\quad \eta(t)\gg 1 ,
    \\[2mm]
    \left|\Psi_{rel}\left(\vec{r}_{ph}\right)\right|^{\,2}\,, &\quad
    \eta(t)\ll 1 ,
    \end{array}
    \right.
\end{equation}
where $\Psi_{rel}$ and $\Psi_{cm}$ are given by
Eqs.~(\ref{ent-free wf}) and (\ref{ent-free wf-2}), but the
arguments of both functions in this case are identical and equal
to $\vec{r}_{ph}$. The width of the single-particle photon wave
packet (\ref{single-w-f}) can be evaluated as
\begin{equation}
    \label{width-single}
    \Delta r_{ph}^{(s)}(t) \sim \max\left\{\frac{c}{\gamma}, \; a(t) \right\}
    \sim \frac{c}{\gamma}\,
    \sqrt{1+\eta^2(t)}\,.
\end{equation}

To describe the single-particle atomic wave packet, we must
integrate Eq.~(\ref{product}) over ${\vec r}_{ph}$ (instead of
${\vec r}_{at}$). In this case, the regions of relatively narrow
and wide atomic proto-packets are separated by the conditions that
the control parameter $\eta(t)$ is either small or large compared
to $v_{rec}/c$ rather than compared to one. This difference with
the case of integration over $\vec{r}_{at}$ is related again to
the form of the entangled argument of the atomic proto-packet in
Eq.~(\ref{product}) discussed previously. The result of
integration over $\vec{r}_{ph}$ is given by
\begin{equation}
    \label{single-wf-at}
    \frac{dw^{(s)}}{d\vec{r}_{at}}=\int d\vec{r}_{ph}\,
    \left|\vec{\Psi}(\vec{r}_{at},\vec{r}_{ph},t)\right|^{\,2}
    \approx\left\{
    \begin{array}{ll}
    \left|\Psi_{cm}(\vec{r}_{at})\right|^{\,2}\,, &\quad \eta(t)\gg v_{rec}/c \\[2mm]
    \left|\Psi_{rel}\left(-\frac{c}{v_{rec}}\,\vec{r}_{at}\right)\right|^{\,2}\,,
    &\quad \eta(t) \ll v_{rec}/c \ .
    \end{array}
    \right.
\end{equation}
The width of this wave packet is evaluated as
\begin{equation}
    \label{width-at-single}
    \Delta r_{at}^{(s)}(t) \sim \max\left\{\frac{v_{rec}}{\gamma}, \; a(t) \right\}
    \sim a(t) \,\frac{\sqrt{\eta^2(t)+v_{rec}^2/c^2}}{\eta(t)}.
\end{equation}

In analogy with Eqs.~(\ref{rel-coinc-ph}) and (\ref{rel-coinc-at})
we can introduce the single-particle relative widths of the photon
and atomic wave packets
\begin{equation}
    \label{rel-single-ph}
    \delta r_{ph}^{(s)}(t) = \frac{\Delta r_{ph}^{(s)}(t)}{c/\gamma}
    \sim \sqrt{1+\eta^2(t)}
\end{equation}
and
\begin{equation}
    \label{rel-single-at}
    \delta r_{at}^{(s)}(t) = \frac{\Delta r_{at}^{(s)}(t)}{a(t)}
    \sim \frac{\sqrt{\eta^2(t)+v_{rec}^2/c^2}}{\eta(t)}.
\end{equation}
By comparing these results with those of Eqs.~(\ref{rel-coinc-ph})
and (\ref{rel-coinc-at}) we find the following group of
fundamental reciprocity relations between the photon vs.~atomic,
and coincidence vs.~single-particle, relative wave packet widths:
\begin{equation}
 \label{symmetry}
 \delta r_{at}^{(c)}(t)\sim\frac{1}{\delta
 r_{ph}^{(s)}(t)}\quad\text{and}\quad \delta r_{ph}^{(c)}(t)
 \sim\frac{1}{\delta r_{at}^{(s)}(t)}.
\end{equation}
Eqs.~(\ref{symmetry}) show that it is possible to find the
coincidence (conditional) wave packet widths without using the
coincidence-scheme measurements. It is sufficient to measure both
single-particle widths $\delta r_{ph}^{(s)}$ and $\delta
r_{at}^{(s)}$, and then the coincidence widths can be found
directly from Eqs.~(\ref{symmetry}) without any further
measurements. More explicitly, we can rewrite
Eqs.~(\ref{symmetry}) into $\Delta r_{at}^{(c)} \Delta
r_{ph}^{(s)} \sim \Delta r_{ph}^{(c)} \Delta r_{at}^{(s)} \sim
a(t) (c/\gamma)$.  Note that $\eta(t) = a(t) / (c/\gamma)$ is
essentially the aspect ratio of the wave packet's $r_{ph} -
r_{at}$ distribution. This definition is rather general: it is
valid also for the momentum widths considered below in Section
VIII, as well as for any other pairs of particles.

\subsection{Entanglement-induced anomalous narrowing and broadening of wave packets}

\begin{figure}[tb]
    \centering\includegraphics[width=16cm]{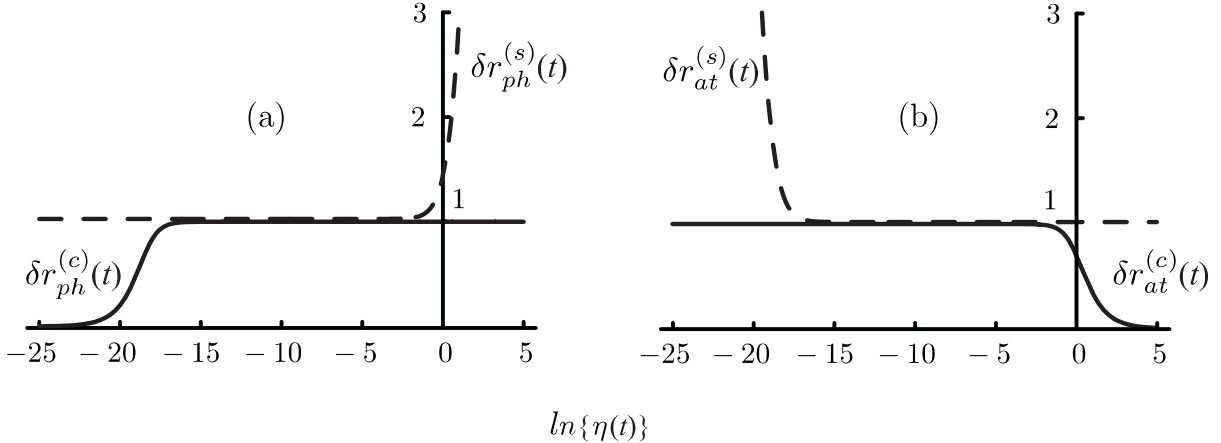}
    \caption{Coincidence (solid lines) and single-particle
    (dashed lines) relative widths of (a) photon and (b) atomic
    wave packets in the coordinate representation.
    We have used $v_{rec}/c=10^{-8}$.}
    \label{Fig4}
\end{figure}

The coincidence and single-particle relative widths of photon wave
packets in the coordinate representation are plotted altogether in
Fig.~\ref{Fig4} in their dependence on the control parameter
$\eta(t)$. The same curves describe atomic and photon wave packets
in the momentum representation, as mentioned below near the end of
Section VIII.
The curves of Fig. 4 show that there are two regions of small and
large values of the parameter $\eta(t)$ where either $\delta
r^{(c)}\ll 1$  or $\delta r^{(s)}\gg 1$. These are the regions of
entanglement-induced narrowing of the coincidence and broadening
of the single-particle wave packets. As is seen from comparison
of Figs. 4(a) and 4(b), for photon and atomic wave packets the
regions of narrowing and broadening are oppositely located.

As an example, let us discuss the physics of these phenomena by
using the photon wave packet widths as shown in
Fig.~\ref{Fig4}(a). In this case the entanglement-induced
narrowing for the coincidence photon wave packet occurs for
$\eta<v_{rec}/c\ll 1$. Qualitatively this effect can be explained
by combining the Doppler effect with the Heisenberg uncertainty
relation~\cite{JHE,JHE1}. According to the latter, the size $a(t)$
of the atomic wave function corresponds to the velocity
uncertainty of the atomic center of mass $\Delta v=1/Ma(t)$. Owing
to the Doppler effect this gives rise to broadening of the
spectrum of emitted photons up to the width $\delta\omega=k\Delta
v=\omega_0/Mca(t)=v_{rec}/a(t)$. If $\eta(t)<v_{rec}/c$, this
broadening exceeds the natural spectral width of the emitted
light, i.e., $\delta\omega>\gamma$. As the photons of all
frequencies are emitted coherently, integration over $\omega$ in
the interval $\delta\omega$ shortens the effective emission time
and spatial size of the emitted photon wave packet down to
$t_{eff}=1/\delta\omega=a(t)/v_{rec}$ and $\Delta
r_{ph}=ct_{eff}=(c/\gamma)(c\eta(t)/v_{rec})$.  They are
respectively much smaller than $\gamma^{-1}$ and $c/\gamma$ if
$\eta(t)\ll v_{rec}/c$. It should be emphasized that the
entanglement-induced wave-packet narrowing can be observed only in
the coincidence scheme of measurements. Fig.~\ref{Fig4}(a) shows
that under the same condition when $\delta r_{ph}^{(c)}<1$ (at
$\eta(t)\ll v_{rec}/c$ or, more specifically,
$ln\{\eta(t)\}\lesssim -18$) the relative width of the
single-particle photon wave packet (the dashed line) remains equal
to one, which corresponds to $\Delta r_{ph}^{(s)}=c/\gamma$.

\begin{figure}[tb]
    \centering\includegraphics[width=13cm]{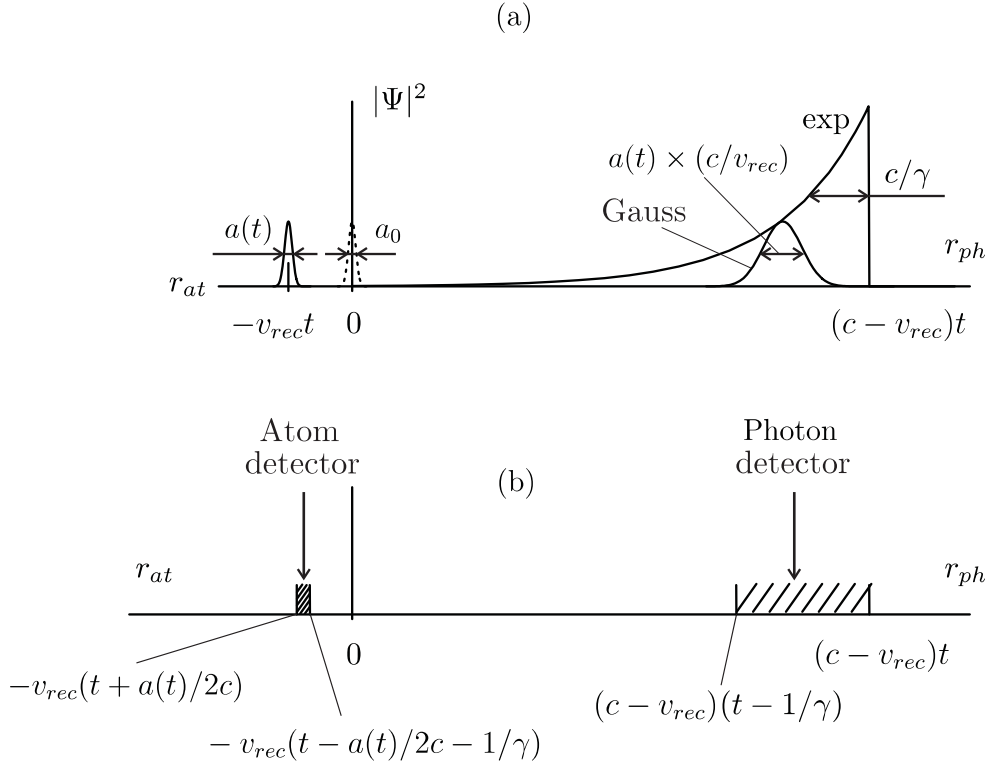}
    \caption{(a) Entanglement-induced narrowing of the photon
    wave packet to be measured in the coincidence scheme and (b) the
    regions where the atom and photon detectors have to be
    installed.}
    \label{Fig5}
\end{figure}

The relation between the coincidence and single particle widths
can also be illustrated by the depiction in Fig.~\ref{Fig5}(a).
The exponential curve with a sharp edge corresponds to the photon
proto-packet $|\Psi_{ph}^{(0)}(\vec{r}_{ph}-\vec{r}_{at})|^2$
given by Eq. (\ref{ent-free wf}) in its dependence on
$\vec{r}_{ph}$. The dotted line is the initial atomic wave packet
$|\Psi_{at}^{(0)}(\vec{r}_{at})|^2$. Two Gaussian solid-line
curves describe the atomic proto-packet
$|\Psi_{at}^{(0)}(\vec{R})|^2$ of Eq. (\ref{ent-free wf-2}) at
$t\neq 0$ in its dependence on $\vec{r}_{at}$ at a given
$\vec{r}_{ph}$ (the left curve) and on $\vec{r}_{ph}$ at a given
$\vec{r}_{at}$ (the right curve). Relative locations of peaks of
these two Gaussian curves are determined by the condition
$\vec{R}=0$, where $\vec{R}$ is defined by Eq. (\ref{cm-argument})
as the argument of the atomic proto-packet (\ref{ent-free wf-2}).
The total two-particle wave function
$\Psi(\vec{r}_{ph},\vec{r}_{at},t)$ of Eq. (\ref{product}) differs
from zero only if the $\vec{r}_{ph}$-dependent exponential and
Gaussian curves overlap with each other. The conditions of their
overlapping are illustrated in Fig.~\ref{Fig5}(b). In this picture
the left shaded area indicates a zone for the atomic detector to
be installed. In this case the $\vec{r}_{ph}$-dependent Gaussian
function overlaps with the exponential one and under the condition
$\eta<v_{rec}/c$ the first of these two curves is narrower than
the second one. The total wave packet, equal to the product of the
Gaussian and exponential functions, has the width of the narrower
one, i.e., of the Gaussian function, and this width is smaller
than than $c/\gamma$. However, if we change the atomic detector
position, the position of the maximum of the Gaussian curve
changes also. The summation of all contributions from various
positions of the atomic detector corresponds to the transition
from coincidence to single-particle measurements and returns us to
the wide exponential curve with the width $c/\gamma$.

\begin{figure}[tb]
    \centering\includegraphics[width=13cm]{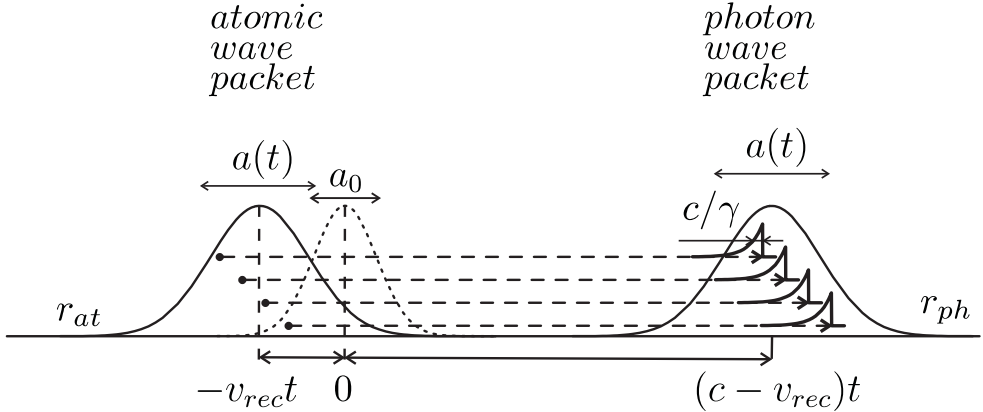}
    \caption{Entanglement-induced broadening of the photon wave packet.}
    \label{Fig6}
\end{figure}

The second new effect illustrated by Figs.~\ref{Fig4}
and~\ref{Fig6} is the entanglement-induced broadening of the
single-particle photon wave packet at large values of the control
parameter, $\eta(t)>1$. Actually, the meaning of the st
Fig.~\ref{Fig6} is practically the same as that of
Fig.~\ref{Fig5}(a). However, the main difference is that the
atomic wave packet is taken very wide now. The dotted line
describes the initial atomic wave packet. Big dots indicate
possible positions of the atomic detector. At any given position
of the atomic detector the photon wave packet to be measured has
an exponential form and the width $c/\gamma$. However, when
contributions from all positions of the atomic detector are summed
together, this gives a wide Gaussian curve for the single-particle
photon wave packet to be observed (the solid-line wide Gaussian
curve at the right-hand side of the picture in Fig.~\ref{Fig6}).
The width of this wide photon wave packet is $\Delta
r_{ph}^{(s)}=a(t)$.

Note that the entanglement-induced broadening of the
single-particle photon wave packet can occur even when the atomic
mass $M$ is taken infinitely large, if only the atomic
center-of-mass wave packet is wide enough, $a(t)>c/\gamma$ or
$\eta(t)> 1$.

\section{Quantification of entanglement and the $R$-parameter}

A convenient measure of the degree of pure-state two-particle
entanglement, calculated in a number of previous
studies~\cite{JHE,JHE1,PRA, LANL,Kazik}, is the Schmidt number $K$
introduced in~\cite{Grobe-etal94},
\begin{equation}
    \label{Schmidt}
    K = \frac{1}{\text{Tr}_{ph}\big(\hat{\rho}^2_{ph}\big)}
    = \frac{1}{\text{Tr}_{at}\big(\hat{\rho}^2_{at}\big)}\,,
\end{equation}
where $\hat{\rho}_{ph}$ and $\hat{\rho}_{at}$ are the reduced
density matrices
\begin{equation}
 \label{reduced}
 \hat{\rho}_{ph}=\text{Tr}_{at}\big(|\Psi\rangle\langle\Psi|\big),\quad
 \hat{\rho}_{at}=\text{Tr}_{ph}\big(|\Psi\rangle\langle\Psi|\big)\,,
\end{equation}
 $\text{Tr}$ denotes trace with respect to either atomic or photon
variables, and $|\Psi\rangle$ is given by
Eq.~(\ref{state-vector}).  Note that $K$ essentially counts the
number of effective Schmidt modes in the Schmidt decomposition of
$\Psi$~\cite{Grobe-etal94}.

\begin{figure}[tb]
    \centering\includegraphics[width=8cm]{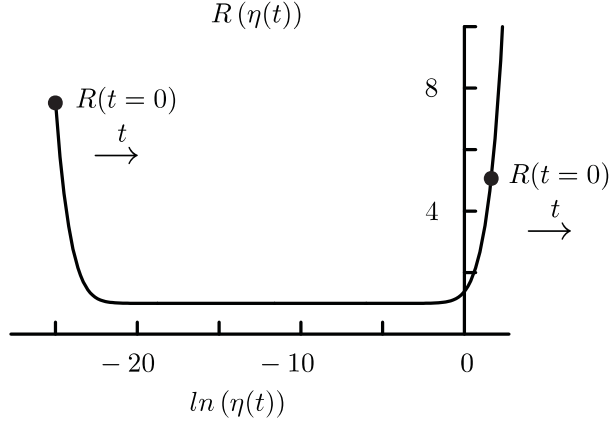}
    \caption{The parameter $R(t)$ vs. the control parameter $\eta(t)$.}
    \label{Fig7}
\end{figure}

On the other hand, by analogy with our earlier discussion of
ionization and dissociation~\cite{PRA}, we can also characterize
the extent of the entanglement that can be seen in coincidence -
single-particle measurements by the ratio $R(t)$ of the
single-particle to coincidence widths of the particles' wave
packets. For spontaneous emission this parameter is given by
\begin{equation}
 \label{ent-param}
 R(t)=\displaystyle\frac{\Delta r_{ph}^{(s)}}{\Delta r_{ph}^{(c)}}
 =\displaystyle\frac{\Delta r_{at}^{(s)}}{\Delta r_{at}^{(c)}}
 = \displaystyle\sqrt{\eta(t)+\frac{1}{\eta(t)}\left(\frac{v_{rec}}{c}\right)^2}
 \times\sqrt{\eta(t)+\frac{1}{\eta(t)}}.
 \end{equation}
The dependence of $R(t)$ on the control parameter $\eta(t)$ is
shown in Fig.~\ref{Fig7}. The parameter $R$ is large at both small
and large values of the control parameter $\eta(t)$, i.e., just
where the above-described entanglement-induced narrowing or
broadening of wave packets occur. At intermediate values of
$\eta(t)$, $R\approx 1$.

Because the atomic center-of-mass wave packet spreads as time
evolves, both $\eta(t)$ and $R(t)$  change with time.  When the
atomic wave packet spreads, $a(t)$ grows with time $t$ whereas the
width of the photon proto-function remains constant and equals
$c/\gamma$. For this reason, in the case of spontaneous emission
the control parameter $\eta(t)$ is always a growing function of
$t$. This contrasts with the case of photoionization~\cite{PRA} in
which, depending on the initial conditions, $\eta(t)$ can be
either an increasing or a decreasing function.  This difference
finds its reflection in the time evolution of the parameter
$R(t)$. The direction of the time evolution of $R(t)$ is shown by
arrows in Fig.~\ref{Fig7} (always to the right!), whereas the dots
indicate two possible initial values of this parameter, $R_0\equiv
R(t=0)$, occurring immediately after the photon emission. In
contrast to this, in the case of photoionization~\cite{PRA}, the
time evolves either to the right (for $R_0$ located at the left
wing of the curve of Fig.~\ref{Fig7}) or to the left (for $R_0$
located at the right wing). The second difference between
spontaneous emission and photoionization concerns the limiting
value of $R$ as $t\rightarrow\infty$. In the case of
photoionization $\eta(\infty)$ is finite and
$R(t\rightarrow\infty)=R_0$ \cite{PRA}. In the case of spontaneous
emission, as the function $\eta(t)$ monotonously grows, the
parameter $R$ grows without limit:
$R\left(\eta(t)\right)\rightarrow\infty$. These differences show
that there is no complete identity between the cases of two
massive particles and a massive plus a massless one.

The relation between the parameter $R$ and entanglement is
dynamical and not direct. In the case of unentangled states
$R\equiv 1$, which means that deviations of $R$ from unity are
related to the entanglement. On the other hand, the parameter $R$
depends on time, whereas in entangled states of non-interacting
particles the degree of entanglement remains constant. This can be
seen clearly in calculations of the Schmidt number $K=const$
\cite{JHE,JHE1,PRA,LANL,Kazik}.  On the other hand, by modelling
the atomic proto-packet of Eq. (\ref{ent-free wf}) by the Gaussian
expressions, in a way similar to that of Ref.~\cite{LANL}, we can
show that $R_0=K$. This means that initially the parameter $R$
coincides with the Schmidt number and shows explicitly the degree
of entanglement of the two-particle state formed immediately after
the decay. Later the parameter $R$ evolves as described above
owing to spreading of the atomic proto-function wave packet. If
initially $R_0>1$ and if at some time the parameter $R(t)$
approaches one, the time region where $R(t)\approx 1$ can be
referred to as the hidden-entanglement region~\cite{LANL}. In this
case the Schmidt number $K$ and the entanglement itself remain as
high as that at $t=0$, but the entanglement cannot be seen or
measured via the comparison of single-particle and coincidence
photon or atomic coordinate wave packet widths.

Here we recall the remarks at the end of Sec. III.  We see that by
substituting $v_{rec}/c$ with $m_e/m_i$, Eq.~(\ref{ent-param})
gives the $R(t)$ of photoionization \cite{PRA}. Now we also see
that, by taking the limit $v_{rec}/c\rightarrow 1$,
Eq.~(\ref{ent-param}) is reduced to that describing the Schmidt
number $K$ for the process of 1D
down-conversion~\cite{Law-Eberly04} (in the two-dimensional case
the down-conversion $K$ equals the square of the one-dimensional
$K$).

\section{Wave packets in the momentum representation}

The widths of photon and atomic wave packets in the momentum
representation is determined by the two-particle atom-photon
momentum wave function determined by Eqs.~(\ref{Gaussian}) and
(\ref{Wigner-W}).
\begin{eqnarray}
    \label{at-phot-mom}
    \Psi^{(mom)}({\vec{q},\,\vec{k}\,})
    &=&
    \left(\displaystyle\frac{L}{2\pi}\right)^3 \, C^{(W-W)}_{\vec{q},\,\vec{k}}
    \nonumber \\
    &\approx&
    -i\displaystyle\frac{e\,z_{eg}\,\omega_0\, a_0^{3/2}\sin\vartheta_{k}}
    {2\pi\sqrt{\omega_{k}}\;\pi^{3/4}}
    \left(-\frac{\vec{q}\cdot\vec{k}}{M}
    +\omega_{k}-\omega_0+\frac{i\gamma}{2}\right)^{-1}
    \exp\left(-\displaystyle\frac{a_0^2({\vec q}+{\vec k})^2}{2}\right).
    \quad
\end{eqnarray}
For analytical calculations it's convenient to substitute the
Lorentzian factor by a Gaussian one
\begin{equation}
    \label{Lorentz-Gauss}
    \Psi^{(mom)}({\vec{q},\,\vec{k}\,})
    \Rightarrow
    -i\displaystyle\frac{e\,z_{eg}\,\omega_0\,a_0^{3/2}\sin\vartheta_{k}}
    {\pi\sqrt{\omega_{k}}\pi^{3/4}\gamma}\,
    \exp\left\{-\frac{1}{2\gamma^2}\left(-\frac{\vec{q}\cdot\vec{k}}{M}
    +\omega_{k}-\omega_0\right)^2 -\displaystyle\frac{a_0^2({\vec q}+{\vec k})^2}{2}\right\}.
\end{equation}
By assuming that the vectors ${\vec q}$ and ${\vec k}$ are
parallel to each other and to the observation direction, we can
find easily from Eq.~(\ref{Lorentz-Gauss}) both coincidence and
single-particle atomic and photon wave packet widths in the
momentum representation:
\begin{eqnarray}
    \label{Delta-q}
    \Delta q^{(c)} &\sim&
    \min\left\{\frac{1}{a_0},\;\frac{\gamma Mc}{\omega_0}\right\}
    \sim \frac{\eta_0}{a_0} \left(\frac{v_{rec}^2}{c^2}+\eta_0^2\right)^{-1/2} \,,
    \nonumber \\
    \Delta q^{(s)} &\sim&
    \max\left\{\frac{1}{a_0},\;\frac{\gamma}{c}\right\}
    \sim \frac{1}{a_0}\sqrt{1+\eta_0^2} \,,
\end{eqnarray}
where $\eta_0$ is the value of the parameter $\eta(t)$ defined in
Eq.~(\ref{eta}) at $t=0$,
\begin{equation}
    \label{eta(0)}
    \eta_0 \equiv \eta(t=0) = \displaystyle\frac{\gamma a_0}{c}\,,
\end{equation}
and
\begin{eqnarray}
    \label{Delta-k}
    \Delta k^{(c)} &\sim&
    \min\left\{\frac{1}{a_0},\;\frac{\gamma}{c}\right\}
    \sim \displaystyle\frac{\gamma/c}{\sqrt{1+\eta_0^2}}\,,
    \nonumber \\
    \Delta k^{(s)} &\sim&
    \max\left\{\frac{\gamma}{c},\;\frac{\omega_0}{Mc^2 a_0}\right\}
    \sim \frac{\gamma/c}{\eta_0}\sqrt{\frac{v_{rec}^2}{c^2}+\eta_0^2}\,
\end{eqnarray}
In contrast to the coordinate wave packet widths the momentum wave
packet widths are independent of time. By introducing the
relative momentum wave packet widths (both coincidence and
single-particle)
\begin{equation}
 \label{rel-q-k}
 \delta q=\frac{\Delta q}{a_0^{-1}},\qquad
 \delta k=\frac{\Delta k}{\gamma/c}\,,
\end{equation}
we find a second group of reciprocity relations
\begin{equation}
 \label{reciprocity}
 \delta k^{(c)}\sim\frac{1}{\delta q^{(s)}}\quad\text{and}\quad
 \delta q^{(c)}\sim\frac{1}{\delta k^{(s)}}.
\end{equation}
By comparing directly the expressions (\ref{Delta-q}) and
(\ref{Delta-k}) for the momentum wave packet widths with
Eqs.~(\ref{rel-coinc-ph}), (\ref{rel-coinc-at}),
(\ref{rel-single-ph}), and (\ref{rel-single-at}) for the
coordinate widths, we find the following series of identities:
\begin{eqnarray}
    \label{identity}
    \delta q^{(c)}\sim\delta r_{ph}^{(c)}(t=0),
    & &
    \delta q^{(s)}\sim\delta r_{ph}^{(s)}(t=0),
    \nonumber \\
    \delta k^{(c)}\sim\delta r_{at}^{(c)}(t=0),
    &\; \text{and} \;&
    \delta k^{(s)}\sim\delta r_{at}^{(s)}(t=0).
\end{eqnarray}
Because of these identity relations the dependencies of the
momentum widths on $\eta_0$ coincide with the dependencies on
$\eta(t)$ of the corresponding coordinate widths. With this
understanding, the reader will be able to see that Fig. \ref{Fig4}
remains just the same in the momentum representation, after
appropriate replacement of the coordinate variances with momentum
variances, and $\eta$ replaced with $\eta_0$ as horizontal axis
label.

Again, as in the case of position-dependent wave packets, Eqs.
(\ref{reciprocity}) can be used  for finding the coincidence
(conditional) momentum-space wave packet widths $\Delta k^{(c)}$
and $\Delta q^{(c)}$ from the single-particle widths $\Delta
k^{(s)}$ and $\Delta q^{(s)}$, without any coincidence-scheme
measurements.

\section{Entanglement and uncertainty relations}

From the identities (\ref{identity}) and reciprocity relations
(\ref{symmetry}) and (\ref{reciprocity}), combined with the
explicit definitions of the parameter $R(t)$ in
Eq.~(\ref{ent-param}) and the wave packet widths in
Eqs.~(\ref{coincidence}), (\ref{at-coinc-width}),
(\ref{width-single}), (\ref{width-at-single}), (\ref{Delta-q}) and
(\ref{Delta-k}), we can find easily the following relation between
the entanglement parameter $K=R(t=0)$ and the widths products
\begin{eqnarray}
    K = R(t=0) &=&
    \displaystyle\sqrt{\eta_0 + \frac{1}{\eta_0} \left(\frac{v_{rec}}{c}\right)^2}
    \times \sqrt{\eta_0 + \frac{1}{\eta_0}}
    \nonumber \\
\label{R-vs-width-product}
    &\sim& \Delta r_{ph}^{(s)}(0) \times \Delta k^{(s)}
    \sim \Delta r_{at}^{(s)}(0) \times \Delta q^{(s)}
    \nonumber \\
    &\sim& \frac{1}{\Delta r_{ph}^{(c)}(0) \times \Delta k^{(c)}}
    \sim \frac{1}{\Delta r_{at}^{(c)}(0) \times \Delta q^{(c)}}
    \gtrsim 1\,.
\end{eqnarray}
The uncertainty relations following from these equations are
\begin{equation}
 \label{Heisenberg}
 \Delta r_{ph}^{(s)}(0)\times\Delta k^{(s)}\sim K\gtrsim  1, \qquad
 \Delta r_{at}^{(s)}(0)\times\Delta
 q^{(s)}\sim K\gtrsim 1
\end{equation}
and
\begin{equation}
 \label{EPR}
 \frac{1}{K}\sim\Delta r_{ph}^{(c)}(0)\times\Delta k^{(c)}\lesssim 1, \qquad
 \frac{1}{K}\sim\Delta r_{at}^{(c)}(0)\times\Delta q^{(c)}\lesssim 1.
\end{equation}
Inequalities~(\ref{Heisenberg}) represent the well-known
Heisenberg uncertainty relations for single-particle measurements
of any particle's coordinate and momentum, while
inequalities~(\ref{EPR}) establish quite different relations
between the particle's conditional coordinate and momentum
uncertainties. These uncertainty relations restrict the products
of such uncertainties from above: the uncertainty products are
equal to (on the order of) the inverse degree of entanglement
$1/K$, and they cannot be larger than one. With a growing degree
of entanglement the products of the conditional uncertainties
fall. On the other hand, Eqs.~(\ref{Heisenberg}) show that the
values of the usual single-particle uncertainty products are equal
to the degree of entanglement $K$. These products grow with a
growing $K$, and they turn to unity only in the non-entangled
states where $K=1$.

Of course, these conclusions do not contradict the usual
Heisenberg inequalities because the coincidence coordinate and
momentum wave packet widths are assumed to be found under
different conditions. With these conditions specified explicitly,
the uncertainty relations (\ref{EPR}) can be written as
\begin{equation}
 \label{EPR-2}
 \Delta r_{ph}^{(c)}(0)|_{{\vec r}_{at}=const}
 \times\Delta k^{(c)}|_{\,{\vec q}=const} \lesssim 1, \qquad
 \Delta r_{at}^{(c)}(0)|_{{\vec r}_{ph}=const}
 \times\Delta q^{(c)}|_{\,{\vec k}=const} \lesssim 1.
\end{equation}
Nevertheless, inequalities (\ref{EPR-2}) determine a kind of law
of nature which, as far as we know, has never been explicitly
formulated. Examples of very small values of the coincidence
uncertainty coordinate and momentum products have been presented
in Refs.~\cite{Kazik, LANL} and experimental observations have
begun to be reported~\cite{Howell}. What is new in the present
derivation is a rather general form of the relationship between
the value of the uncertainty products and the degree of
entanglement.

In some sense the conditional uncertainty relations~(\ref{EPR})
are in concord with the well-known (often termed paradoxical)
prediction by Einstein, Podolsky, and Rosen \cite{EPR}. Indeed, as
it was shown in~\cite{EPR}, in entangled decaying bipartite
systems momentum or coordinates of one particle can be measured
more precisely than permitted by the Heisenberg uncertainty
relation if \emph{prior to this} appropriate measurements are done
\emph{exclusively} with the other particle. But, of course, in
such a formulation there is a rather well pronounced difference
with our approach. The conditional uncertainty
relations~(\ref{EPR}) are based on the idea of coincidence
measurements, i.e., \emph{simultaneous} and \emph{joint}
measurements to be made with both particles.

Eqs.~(\ref{EPR}) and (\ref{EPR-2}) are derived at $t=0$, and they
are valid as long as the atomic wave packet does not significantly
spread. It is interesting to check how do the derived relations
change at longer $t$ or, in other words, whether the wave packet
spreading modifies or violates the uncertainty relations
(\ref{EPR}) and (\ref{EPR-2}). The general answer is no, there is
no violation of the time-dependent coincidence uncertainty
relations arising from the wave packet spreading. To show this we
have to use explicit expressions for the time-dependent widths of
the coordinate wave packets $\Delta r_{ph}^{(c)}(t)$ and $\Delta
r_{at}^{(c)}(t)$ given in Eqs.~(\ref{coincidence})
and~(\ref{at-coinc-width}) together with Eqs.~(\ref{Delta-k})
and~(\ref{Delta-q}) for $\Delta k^{(c),(s)}$ and $\Delta
k^{(c),(s)}$. The results have the form
\begin{eqnarray}
    \label{EPR(t)-ph}
    \Delta r_{ph}^{(c)}(t) \times \Delta k^{(c)}
    \sim \frac{\eta(t)}{\sqrt{\eta^2(t) + v_{rec}^2/c^2}}
    \times \frac{1}{\sqrt{\eta_0^2 + 1}}
    = \frac{1}{R(t)} \sqrt{\frac{\eta^{2}(t)+1}{\eta_0^2 + 1}}
\end{eqnarray}
and
\begin{equation}
    \label{EPR(t)-at}
    \Delta r_{at}^{(c)}(t) \times \Delta q^{(c)}
   \sim \frac{1}{\sqrt{1+\eta^2(t)}} \times
    \frac{\eta_0}{\sqrt{\eta_0^2 + v_{rec}^2/c^2}}.
\end{equation}
 As $\eta(t)\geq\eta_0$, Eqs. (\ref{EPR(t)-ph}) and
(\ref{EPR(t)-at}) show clearly that, indeed, the conditional
uncertainty relations (\ref{EPR-2}) remain valid at any $t$,
\begin{equation}
 \label{EPR-3}
 \Delta r_{ph}^{(c)}(t)\times\Delta k^{(c)} \lesssim 1, \qquad
 \Delta r_{at}^{(c)}(t)\times\Delta q^{(c)} \lesssim 1.
\end{equation}
But Eqs.~(\ref{EPR(t)-ph}) and~(\ref{EPR(t)-at}) indicate also
some difference between the coincidence uncertainty products for
atoms and photons in their dependence on time $t$. If the photon
uncertainty coincidence product (\ref{EPR(t)-ph}) monotonously
grows with a growing $t$, though remaining smaller than one, the
atomic coincidence uncertainty product~(\ref{EPR(t)-at})
monotonously falls approaching zero at very large values of
$\eta(t)$. This asymmetry is related to the zero photon mass and
non-zero mass of an atom, which results in spreading atomic and
non-spreading photon proto-functions (given,correspondingly by
Eqs. (\ref{ent-free wf-2}) and (\ref{ent-free wf})).

What is destroyed by spreading is a direct connection between
conditional uncertainty products and the Schmidt number $K$, at
$t=0$ given by Eq. (\ref{EPR}). As $K=const$ and the photon
conditional uncertainty product (\ref{EPR(t)-ph}) is a growing
function of $t$, owing to spreading this product becomes larger
than $1/K$. On the other hand, the last form of Eq.
(\ref{EPR(t)-ph}) shows that this product is also larger than
$1/R(t)$. Hence, at $t\neq 0$ the photon conditional uncertainty
product appears to be restricted from both above and below
\begin{equation}
 \label{EPR(t)-ph-2}
 \max\left\{\frac{1}{K},\,\frac{1}{R(t)}\right\}<\Delta r_{ph}^{(c)}(t)\times\Delta k^{(c)} <
 1.
\end{equation}
As for the atomic conditional uncertainty product, as it falls
with a growing $t$, owing to spreading it becomes even smaller
than $1/K$, and this restricts this product from above even
stronger than without spreading (see Eq. (\ref{EPR-3}))
\begin{equation}
    \label{EPR(t)-at-2}
    \Delta r_{at}^{(c)}(t) \times \Delta q^{(c)}
    \lesssim\frac{1}{K}\lesssim 1.
\end{equation}

\section{Experiment}

Concerning related experiments, note first the two recent works
~\cite{Howell,Kulik} on investigation of entanglement in the
process of spontaneous parametric downconversion. We did not
consider down-conversion in this paper at all.  On the other hand,
many results presented above are general enough to be valid for
any pairs of particles. Definitely, one of such results is the
relation between the degree of entanglement and the value of the
product of coordinate and momentum conditional uncertainties
(\ref{EPR}). In the experiment ~\cite{Howell} this product was
found to be about 0.2 which is clearly less than one. This result
agrees with Eqs.~(\ref{EPR}) derived above, though the relation
between the conditional uncertainty product and the Schmidt number
$K$ or the wave packet width ratio $R$ were not checked
experimentally. We think that such an additional experimental
investigation would be very interesting.

As for experiments specifically on entanglement in spontaneous
emission of a photon, the most closely related work is that by
Kurtsiefer, et al.~\cite{Kurt}. In this experiment the atomic
momentum wave packet arising after spontaneous emission of a
photon was measured in the coincidence and single-particle schemes
of measurements. The coincidence width was shown to be smaller
than the single-particle one. But this is not yet a direct
confirmation of our predictions. As shown above, at small values
of the control parameter $\eta$ the coincidence (conditional)
width of the atomic momentum wave packet $\Delta q^{(c)}$ falls
below its natural value $a_0^{-1}$ equal to the same width for a
non-emitting ground-state atom (Fig.~\ref{Fig2}(a)). To see this
one has to provide conditions under which $\eta<v_{rec}/c$.
Probably this requirement was not fulfilled in the
experiment~\cite{Kurt} and for this reason the observed
coincidence wave packet width remained larger than $a_0^{-1}$.
This quick analysis shows that, first, observation of entanglement
effects in atomic spontaneous emission is possible and, second, to
observe these effects the experiment has to be further refined.

\section{Conclusion}

To summarize, the relationship between atom-photon wave packet
structures and entanglement in spontaneous emission has been
analyzed. Finite atomic mass and finite initial size of the atomic
center-of-mass wave function were taken into account.  Two new
effects reported were anomalous narrowing and anomalous broadening
of the coordinate atomic and photon wave packets as observed in
the coincidence and single-particle schemes of measurements.
Atomic and photon wave packets were investigated both in the
coordinate and momentum representations and a series of symmetry
relations for their widths were established. These relations and
the definition of the parameters characterizing the degree of
entanglement were used to establish that (a) the product of
single-particle coordinate and momentum uncertainties is equal (or
is of the order of) the Schmidt number $K$ and (b) the product of
coincidence (conditional) coordinate and momentum uncertainties is
equal or is of the order of the inverse Schmidt number, $1/K$. The
second of these two results shows that the coincidence coordinate
and momentum uncertainty product is always less than one, and this
is in the spirit of the conclusion reached by EPR in their famous
discussion.

\acknowledgements The research reported here has been supported
under the NSF grant PHY-0072359, Hong Kong Research Grants Council
(grant no. CUHK4016/03P), the RFBR grants 02-02-16400 and
05-02-16469, the Russian Science Support Foundation (MAE), and the
award of a Messersmith Fellowship (KWC).


\end{document}